\documentclass[twocolumn,english]{revtex4}
\usepackage[T1]{fontenc}
\usepackage[latin9]{inputenc}
\usepackage{graphicx}
\usepackage{esint}

\makeatletter

\DeclareRobustCommand{\cyrtext}{%
  \fontencoding{T2A}\selectfont\def\encodingdefault{T2A}}
\DeclareRobustCommand{\textcyr}[1]{\leavevmode{\cyrtext #1}}
\AtBeginDocument{\DeclareFontEncoding{T2A}{}{}}

\providecommand{\tabularnewline}{\\}

\@ifundefined{textcolor}{}
{%
 \definecolor{BLACK}{gray}{0}
 \definecolor{WHITE}{gray}{1}
 \definecolor{RED}{rgb}{1,0,0}
 \definecolor{GREEN}{rgb}{0,1,0}
 \definecolor{BLUE}{rgb}{0,0,1}
 \definecolor{CYAN}{cmyk}{1,0,0,0}
 \definecolor{MAGENTA}{cmyk}{0,1,0,0}
 \definecolor{YELLOW}{cmyk}{0,0,1,0}
 }

\makeatother

\usepackage{babel}

\begin{document}

\title{$O(N)$ continuous electrostatics solvation energies calculation
method for biomolecules simulations.}

\author{P.O. Fedichev$^{1}$, E.G. Getmantsev$^{1}$, L.I. Menshikov$^{2}$}

\affiliation{$^{1)}$Quantum Pharmaceuticals Ltd, Ul. Kosmonavta Volkova 6\textcyr{\char192}-606,
Moscow, Russia }

\email{peter.fedichev@q-pharm.com}

\homepage{http://www.q-pharm.com}

\address{$^{2)}$RRC Kurchatov Institute, Kurchatov Square 1, 123182, Moscow,
Russian Federation }
\begin{abstract}
We report a development of a new fast surface-based method for numerical
calculations of solvation energy of biomolecules with a large number
of charged groups. The procedure scales linearly with the system size
both in time and memory requirements, is only a few percents wrong
for any molecular configurations of arbitrary sizes, gives explicit
value for the reaction field potential at any point, provides both
the solvation energy and its derivatives suitable for Molecular Dynamics
simulations. The method works well both for large and small molecules
and thus gives stable energy differences for quantities such as solvation
energies of molecular complex formation. 
\end{abstract}
\maketitle
Solvent interactions play an essential role in Nature in determining
electrostatic potential energies of molecular conformations, charge
states of proteins, dissociation constants of small molecules, and
binding properties of protein-ligand complexes. A solvation energy
calculation for a molecule-sized object has always been and still
is a challenging problem. The most accurate approach is, apparently,
a large scale MD simulation \cite{rapaport2004amd,ModernMD} of the
body of interest immersed in a tank of water molecules in a realistic
force field or even within quantum mechanical settings. Though such
an approach may in principle provide ultimately accurate predictions,
the calculations are time consuming and pose a number of specific
problems stemming, e.g. from long relaxation times of water clusters.
One possible way to bridge the {}``simulation gap'' is to employ
different types of continuous solvation models. Fortunately, water
is characterized by a very large value of dielectric constant and
therefore the reaction field of water molecules is collective in nature.
Although realistic properties of molecular interactions depend both
on short-scale water molecules alignment and on the long-range dipole-dipole
interactions at the same time \cite{fedichev2008fep,fedichev2006long},
purely electrostatic models, such as Poisson-Boltzmann equation solvers
\cite{baker2001ena,schafer2000cit}, turned out to be very successful
in various applications. 

Even within the realm of continuous electrostatic models there are
numerous approaches in use to calculate the polar part of the solvation
energies. Popular techniques span from finite element methods (FEM,
\cite{baker2001ena,schafer2000cit,bordner2003bes,boschitsch2002fbe,boschitsch2004hbe,horvath1996dap,vorobjev1997fam,zhou1993bes})
to various types of Generalized Born (GB) approximations \cite{schaefer1996cat,still1990sts,tsui2001taa,simonson2001mec,lee2002ngb,hassan2002cac,rashin1990hpc,feig2004pcg,onufriev2002ebr}.
A numerical FEM solution to the Poisson-Boltzmann equation (PBE) is
a formally fast (the calculation time and memory scale $\propto N,$
with $N$ being the number of particles in the system) and is a rigorous
attempt to solve the electrostatics problem. On the other hand FEM
involve a good numerical overhead and in practice GB approximations
are faster, in spite of the fact that it normally takes $O(N^{2})$
operations to calculate GB energy. Unfortunately GB approximations
are very rough and that is why GB calculations work well only for
small and medium sized molecules, whereas FEM methods can, although
at expense of a numerical complexity, be applied to very large systems.
The particular boundary between the applicability of the two methods
depends on the balance of speed the amount of details and accuracy
required in a specific application.

In this Letter we push forward our recently established connection
\cite{fedichev2009fast} between the Generalized Born (GB) models
\cite{still1990sts,tsui2001taa,lee2002ngb,bashford2000geb,onufriev2002ebr,lee2003new,mongan2007analysis}
and the boundary integral formulation of the electrostatics problem
\cite{schafer2000cit}. We show that the GB solvation energy can in
fact be calculated in linear time and memory for arbitrary system
of charges. Using post-Coulomb Field Approximation (post-CFA) for
the Born radii calculations we report a development of a new fast
surface-charges density based method (Surface Charges GB, SCGB) for
numerical solution of the Poisson-Boltzmann equations and use it for
the solvation energy calculations for biomolecules with a large number
of charged groups. The procedure turns out only a few percents wrong
for realistic molecular configurations, gives explicit value for the
reaction field potential at any point within the system and provides
both the solvation energy and its derivatives suitable for Molecular
Dynamics (MD) simulations. The method works well both for large and
small molecules and thus gives stable energy differences for quantities
such as solvation energies of a molecular complex formation.

\begin{figure}
\includegraphics[width=0.9\columnwidth]{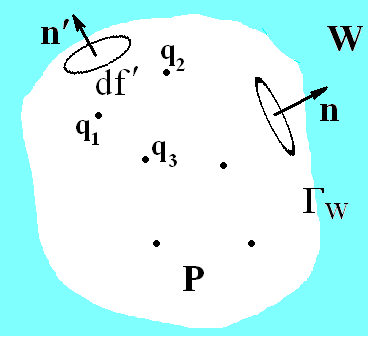}

\caption{Schematic representation of a charged biomolecule in a continuous
water model.\label{fig:Biomolecule-in-water.}}

\end{figure}

Modern implicit water methods trade accuracy and physical sophistication
for speed and usually are based on assumptions \cite{lee2002ngb,mongan2007analysis}
traceable back to the original approach of Born \cite{born1920volumes}.
Consider a molecule modeled as a system of charges confined within
a water cavity as shown on Fig.\ref{fig:Biomolecule-in-water.}. The
shape of the cavity can be either obtained by displacing the water
out of all the atomic volumes and then collecting the atomic volumes
into the molecular volume \cite{lee2002ngb,lee2003new,tjong2007gbr6}.
An alternative can be the molecular volume separated from the water
bulk by a sufficiently smooth interface surface $\Gamma_{W}$ containing
all the atoms and having no unphysical water-filled caverns inside
\cite{lee1971interpretation,richards1977areas}. Neither of approaches
is ideal, though the surface based methods often produce better molecular
volumes. The polar contribution to the solvation energy (the solvation
energy) of the molecule is given by

\begin{equation}
E_{S}=\frac{1}{2}\sum_{i=1}^{N}q_{i}\varphi_{1}(\mathbf{r}_{i}),\label{eq: Solvation energy}\end{equation}
where $\varphi_{1}$ is the so called reaction field potential produced
by the water polarization charges as explained in e.g. \cite{schafer2000cit}.
Here the Latin indices $i=1...N$ enumerate the charges, $N$ is the
total number of charges, $q_{i}$ and $\mathbf{r}_{i}$ are the charges
and the positions of the ions. 

The actual calculation of the reaction field potential depends on
further assumptions and may be performed in a number of ways. Since
the dielectric constant in water is large ($\epsilon_{W}\approx80\gg1$),
the electric potential on the molecular surface vanishes to a very
good accuracy: $\varphi(\mathbf{r})\mid_{\Gamma_{W}}=0$ (the so called
{}``ideal conductor'' approximation). Therefore the polarization
charges are confined to the interface and the reaction field can be
approximated as \begin{equation}
\varphi_{1}(\mathbf{r})=\int_{\Gamma_{W}}df^{\prime}\frac{\sigma_{S}\left(\mathbf{r}^{\prime}\right)}{\left|\mathbf{r}-\mathbf{r}^{\prime}\right|}.\label{eq: phi 1 exact}\end{equation}
Here $\sigma_{S}\left(\mathbf{r}^{\prime}\right)$ is the surface
density of the polarization charges and $df$ is the molecular surface
element. The total electric potential at a given point $\mathbf{r}$
is $\varphi(\mathbf{r})=\varphi_{0}(\mathbf{r})+\varphi_{1}(\mathbf{r}),$
where \[
\varphi_{0}(\mathbf{r})=\sum_{j=1}^{N}\frac{q_{j}}{\left|\mathbf{r}-\mathbf{r}_{j}\right|}\]
is the Coulomb potential generated by the molecular charges. The surface
charge density $\sigma$ satisfies the integral equation\begin{equation}
2\pi\sigma\left(\mathbf{r}\right)+\int_{\Gamma_{W}}df^{\prime}\sigma\left(\mathbf{r}^{\prime}\right)\frac{\mathbf{n\left(\mathbf{r}-\mathbf{r}^{\prime}\right)}}{\left|\mathbf{r}-\mathbf{r}^{\prime}\right|^{3}}=-\sum_{i}q_{i}\frac{\mathbf{n}\left(\mathbf{r}-\mathbf{r}_{i}\right)}{\left|\mathbf{r}-\mathbf{r}_{i}\right|^{3}}.\label{eq: Eq for sigma}\end{equation}
If the molecular surface is properly discreticized then both the polarization
charge density $\sigma$ and the solvation energy can be obtained
iteratively in $O(NlnN)$ operations with the help of either FFT or
fast multipole methods for fast matrix-vector products and proper
preconditioners \cite{greengard1987fast,sagui1999molecular,beckers1998iterative,de1980simulation,essmann1995smooth}.
In practice the number of iterations required for full convergence
is far from a few and the whole calculation is nevertheless fairly
computationally demanding. Another problem arises from the fact that
applications such as MD simulations or minimizations require derivatives
with respect to the atoms coordinates. Naturally, finding a derivative
of an iteratively obtained solution is not an easy task. That is why
a substantial effort was put in finding reasonable approximate solutions
to Eq. (\ref{eq: Eq for sigma}) as described in the recent publications
\cite{bardhan2008interpreting,bardhan2009bounding} and the refs.
therein.

Historically Generalized Born (GB) methods provide an apparently different
way of the solvation energy calculations \cite{schaefer1996cat,still1990sts,tsui2001taa,simonson2001mec,lee2002ngb,hassan2002cac,rashin1990hpc,feig2004pcg,onufriev2002ebr}.
In our recent work \cite{fedichev2009fast} we established the link
between the surface electrostatics and GB models. It turns out that
GB models can be also used to provide the reaction field potential
approximation within the molecule and to calculate the polarization
charge density. To do that we reintroduce GB models following our
presentation in \cite{fedichev2009fast} using the simplest Kirkwood-like
form of the reaction field potential \cite{Kirkwood,grycuk2003deficiency}

\begin{equation}
\varphi(\mathbf{r})=\varphi_{0}+\varphi_{1}\approx\sum_{j=1}^{N}q_{j}\left(\frac{1}{\left|\mathbf{r}-\mathbf{r}_{j}\right|}-\frac{1}{S_{j}}\right),\label{eq: phi in the kirkwood approximation}\end{equation}
where \[
S_{j}(\mathbf{r})=\sqrt{\left(\mathbf{r}-\mathbf{r}_{j}\right)^{2}+R(\mathbf{r}_{j})R\left(\mathbf{r}\right)},\]
and $R(\mathbf{r})$ is a properly chosen function. Specific expressions
for the function $R$ are different in different models and are expressed
either in terms of either volume \cite{schaefer1996cat,still1990sts,tsui2001taa,simonson2001mec,lee2002ngb,hassan2002cac,rashin1990hpc,feig2004pcg,onufriev2002ebr}
or surface integrals \cite{Onufriev,lee2003new,lee2002ngb,tjong2007gbr6}.
The values of the function $R(\mathbf{r})$ at the positions of the
charges are called the respective Born radii, $R_{i}=R(\mathbf{r}_{i})$
of the ions. The surface and the volume integral formulation dichotomy
of GB models has a long history and the models defined with a help
of properly chosen molecular surfaces (see e.g. \cite{lee1971interpretation,richards1977areas})
have a good number of practical advantages \cite{romanov2004surface}.
Normally the polar part of the solvation energy is obtained by plugging
the expression from the Eq. (\ref{eq: phi in the kirkwood approximation})
for the reaction field potential into the Eq. (\ref{eq: Solvation energy})
\cite{Kirkwood,grycuk2003deficiency,fedichev2009fast}: \begin{equation}
\left(E_{S}\right)_{GB}=-\frac{1}{2}\sum_{i,j=1}^{N}\frac{q_{i}q_{j}}{f\left(r_{ij}\right)}\left(\frac{1}{\epsilon_{P}}-\frac{1}{\epsilon_{W}}\right),\label{eq: New expression for solvation energy}\end{equation}
where $f(r_{ij})=S_{j}(\mathbf{r}_{i})$, $\epsilon_{P}\sim1$ is
the dielectric constant of the molecular interior. The expression
implies double summation over the molecular charges and requires $O(N^{2})$
operations to compute. 

One of the simplest way to calculate the Born radii comes from the
so called Coulomb field approximation (CFA) \cite{schaefer1996cat,gilson2007calculation,schlick1999algorithmic,bardhan2008interpreting,bardhan2009bounding}:
the electric displacement vector in a nonuniform medium is taken as
that in the vacuum. The CFA is wrong for the ions next to the protein
boundary \cite{lee2002ngb,lee2003new,mongan2007analysis,grycuk2003deficiency},
which is a problem indeed, since most of the charges within typical
biomolecules are located next to the molecular surfaces. There are
a few ways to go beyond the CFA and obtain a more accurate approximation.
The first class of the models was introduces in a number of works
\cite{lee2002ngb,mongan2007analysis,lee2003new,grycuk2003deficiency,fedichev2009fast,romanov2004surface,ghosh1998generalized,sigalov2005incorporating}
in either of the equivalent volume or surface integral formulations
\begin{equation}
\frac{1}{R_{i}^{\beta-3}}=\frac{\beta-3}{4\pi}\int_{W}\frac{d^{3}r^{\prime}}{s_{i}^{\beta}}=\frac{1}{4\pi}\int_{\Gamma_{W}}\frac{\left(\mathbf{n^{\prime}}\mathbf{s}_{i}\right)}{s_{i}^{\beta}}df^{\prime},\label{eq: General expression}\end{equation}
where $s_{i}=\left|\mathbf{s}_{i}\right|$, $\mathbf{s}_{i}=\mathbf{r}^{\prime}-\mathbf{r}_{i}$,
and $\beta=5-7$ is a (variational) parameter. The integration over
the water bulk $W$ in the middle of (\ref{eq: General expression})
is transformed to the equivalent boundary integral form in a standard
way with the help of the Gauss theorem \cite{ghosh1998generalized}.
Another important model is given by \begin{equation}
\frac{1}{R_{i}^{\alpha-2}}=C_{\alpha}\int_{\Gamma_{w}}\frac{df^{\prime}}{|\mathbf{r}_{i}-\mathbf{r}^{\prime}|^{\alpha}},\label{eq: Surface Born for different alpha}\end{equation}
where $C_{\alpha}$ is the properly chosen constant. The special choice
of $\beta=6$ in the model (\ref{eq: General expression}) and the
two models described by Eq. (\ref{eq: Surface Born for different alpha})
with $\alpha=3,4$ and $C_{\alpha}=1/4\pi$ are exact for an arbitrary
system of charges within a spherical {}``molecule'' \cite{fedichev2009fast}.
Though the specific choice of the Born radii method is not important
for the following considerations, we naturally prefer these {}``inherently
accurate'' models and call them SCGB (Eq. (\ref{eq: General expression})
with $\beta=6$) and SCGB(3) or SCGB(4) (Eq. (\ref{eq: Surface Born for different alpha})
with $\alpha=3,4$).

To obtain a faster method we suggest to use the model potential (\ref{eq: phi in the kirkwood approximation})
to calculate the polarization charge density on the molecular surface
$\sigma$ from the electrostatic boundary condition in a standard
way \begin{equation}
\sigma=\frac{1}{4\pi}\frac{\partial\varphi}{\partial n}.\label{eq: Exact formula for sigma_S}\end{equation}
Next to the molecular surface ($\mathbf{r}^{\prime}\rightarrow\Gamma_{W}$)
the functions $R$ in each of our models vanishes, $R\left(\mathbf{r}^{\prime}\right)\approx2h\rightarrow0$,
where $h$ is the distance from a given point to the surface \cite{fedichev2009fast}.
Therefore \begin{equation}
\sigma\left(\mathbf{r}^{\prime}\right)\approx-\frac{1}{4\pi}\sum_{j}q_{j}\frac{R_{j}}{\left|\mathbf{r}^{\prime}-\mathbf{r}_{j}\right|^{3}}.\label{eq: Sigma_S in FSBE approximation}\end{equation}
Once the surface charge density is known, we can use Eq. (\ref{eq: phi 1 exact})
to obtain the solvation energy in the same way as if $\sigma$ is
a solution of the surface electrostatics. In what follows we call
the method Surface Charges Generalized Born (SCGB).

Let us summarize the solvation energy calculation algorithm in a few
lines:
\begin{enumerate}
\item given a set of molecular charges $q_{i}$ located at the positions
$\mathbf{r}_{i}$ and a useful discretization of the surface, representing
the molecule-water interface, we calculate first the set of Born radii
with the help of the surface integration according to either of Eq.
(\ref{eq: General expression}) and Eq. (\ref{eq: Surface Born for different alpha})
with properly chosen values of $\alpha$ or $\beta$. 
\item as soon as the Born radii are ready, we calculate the surface charge
density at every point on the molecular surface according to Eq.(\ref{eq: Sigma_S in FSBE approximation}).
\item now when the surface charge density is known, we can calculate the
solvation energy using the exact expressions (\ref{eq: Solvation energy})
and (\ref{eq: phi 1 exact}). 
\end{enumerate}
Although the apparent computational complexity of the outlined procedures
is $O(M\times N)$ (or better say in $O(M\times N\log(M\times N))$),
where $M$ is the number of points in the discrete representation
of the molecular surfaces, the discrete summation involves only the
coordinates differences and thus the calculation can be performed
in $O(M+N)$ operations with the help of either FFT or fast multipole
methods.

SCGB approximation is by no means exact, $\triangle\varphi\neq0$,
and hence there can be (and in fact there are) superficial polarization
charges in the water bulk and within the molecule. Let us perform
a few simple model calculation to see how accurate the suggested SCGB
procedure can be. Consider first a charge placed somewhere at the
position $\mathbf{r}_{j}$ within a sphere of a radius $a$. Then,
a simple calculation yields\begin{equation}
R_{j}=\frac{a^{2}-r_{j}^{2}}{a}\label{eq: Modified Born radius for sphere}\end{equation}
and the model expression for the electrostatic potential coincide
with the exact result e.g. from \cite{landau1961electrodynamics}\begin{equation}
\varphi(\mathbf{r})=q_{j}\left[\frac{1}{\left|\mathbf{r}-\mathbf{r}_{j}\right|}-\frac{a}{\left|r_{j}\mathbf{r}-a^{2}\widehat{\mathbf{r}}_{j}\right|}\right],\label{eq: Exact potential for ion in sphere}\end{equation}
with $\widehat{\mathbf{r}}_{j}=\mathbf{r}_{j}/r_{j}$. In the same
way the surface charge density calculated from this expression for
the potential according to Eq.(\ref{eq: Exact formula for sigma_S})
coincides with that given by Eq. (\ref{eq: Sigma_S in FSBE approximation}):\[
\sigma_{j}=-\frac{a^{2}-r_{j}^{2}}{4\pi a\left|\mathbf{r}^{\prime}-\mathbf{r}_{j}\right|^{3}}.\]
Since $\sigma_{S}=\sum_{j}\sigma_{j}$ is an additive quantity, SCGB
approximations gives the exact result for $\sigma_{S}$ for an arbitrary
charge distribution within a sphere. An interesting case corresponds
to a sphere with $a=\infty$, that is a very large molecule occupying
a half-space. 

\begin{figure}
\includegraphics[width=0.9\columnwidth]{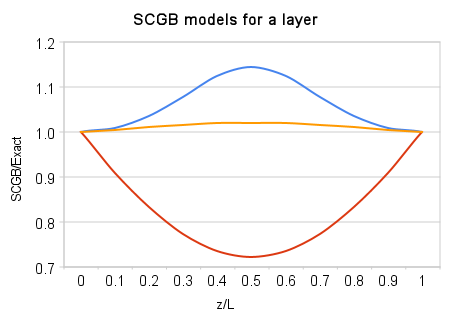}

\caption{Solvation energies to exact solvation energy ratios comparison for
a charge placed within a dielectric layer of thickness $L$. The upper,
middle, and lower curves describe the solvation energy for $SCGB$,
$SCGB(3,4)$, correspondingly. All the quantities approach the exact
value on the molecule boundaries ($z=0,L$). \label{fig:Ratios-the-results} }

\end{figure}

SCGB approach can not, of course, be exact for an arbitrary molecule
geometry.  Consider another practically important example: a plain
layer-like {}``molecule'' (or membrane) of the thickness $L$ surrounded
by the continuous water on both sides with a charge $q$ placed inside
the layer at the distance $z$ from one of the water interface planes.
The exact result for the solvation energy of the system is \cite{Stratton,jackson1999ce}\begin{equation}
\left(E_{S}\right)_{ex}=\frac{q^{2}}{L}\int_{0}^{\infty}dk\left[\frac{\sinh\left(k\bar{z}\right)\sinh\left(k\left(1-\bar{z}\right)\right)}{\sinh\left(k\right)}-\frac{1}{2}\right],\label{eq: Exact solvation energy for the layer}\end{equation}
where $\bar{z}=z/L$. The results for the layer-like molecule in all
the three SCGB approaches are:

\[
\left(E_{S}\right)_{SCGB}=-\frac{q^{2}}{L}\frac{1-2\overline{z}\left(1-\overline{z}\right)}{4\overline{z}\left(1-\overline{z}\right)\sqrt[3]{1-3\overline{z}\left(1-\overline{z}\right)}},\]

\[
\left(E_{S}\right)_{3}=-\frac{q^{2}}{L}\frac{1-2\overline{z}\left(1-\overline{z}\right)}{4\overline{z}\left(1-\overline{z}\right)},\]
\[
\left(E_{S}\right)_{4}=-\frac{q^{2}}{L}\frac{\sqrt{1-2\overline{z}\left(1-\overline{z}\right)}}{4\overline{z}\left(1-\overline{z}\right)}.\]
We compared them with the exact result of Eq. (\ref{eq: Exact solvation energy for the layer})
on Figure \ref{fig:Ratios-the-results}. All the quantities approach
the exact value on the molecule boundaries ($z=0,L$) and differ from
the exact solution in the middle of the layer.

Another challenging case is the calculation for a single charge $q$
placed within a wedge made of the two perpendicular infinite walls
(the {}``$xz$'' and {}``$yz$'' planes). The SCGB results are

\[
\left(E_{S}\right)_{SCGB}=-\frac{q^{2}}{r}\frac{\left(2-\sin\varphi\cos\varphi\right)\left(\sin\varphi+\cos\varphi\right)}{8\sin\varphi\cos\varphi\sqrt[3]{1-\frac{3}{2}\left(\sin\varphi\cos\varphi\right)^{2}}},\]
\[
\left(E_{S}\right)_{3}=-\frac{q^{2}}{r}\frac{\pi\left(2-\sin\varphi\cos\varphi\right)\left(\sin\varphi+\cos\varphi\right)}{8\sin\varphi\cos\varphi\left[\left(\pi-\varphi\right)\cos\varphi+\left(\frac{\pi}{2}+\varphi\right)\sin\varphi\right]},\]
\[
\left(E_{S}\right)_{4}=-\frac{q^{2}}{r}\frac{\sqrt{\left(2-\sin\varphi\cos\varphi\right)\left(\sin\varphi+\cos\varphi\right)}}{4\sqrt{2}\sin\varphi\cos\varphi},\]
where $\varphi$ is the azimuthal angle between the position of a
charge and the $xz$-plane, $r$ is the distance separating the charge
from the wedge. The results should be compared with the exact solvation
energy \[
\left(E_{S}\right)_{ex}=-\frac{q^{2}}{r}\frac{\sin\varphi+\cos\varphi-\sin\varphi\cos\varphi}{4\sin\varphi\cos\varphi}.\]
The error can be analyzed by observing the ratio $Q_{a}=\left(E_{S}\right)_{a}/\left(E_{S}\right)_{ex}$,
which is the largest at $\varphi=\pi/4$ and \[
Q_{SCGB}=1.36,\; Q_{3}=0.91,\; Q_{4}=1.13.\]
The measures of the error are reasonable though of course not perfect.
To build more confidence we have also performed the calculations for
a charge placed on the axis of an infinite cylinder of the radius
$R$ and in the center of a cube of the size $a$ (see Table \ref{tab:SCGB-calculations-examples})
with roughly the same results.

\begin{table}
\begin{tabular}{|c|c|c|c|c|}
\hline 
 & Exact & SCGB/Ex & SCGB(3)/EX & SCGB(4)/EX\tabularnewline
\hline
\hline 
cylinder & $-0.436q^{2}/R$ & $0.75$ & $0.90$ & $1.02$\tabularnewline
\hline 
cube & $-0.874q^{2}/a$ & $1.13$ & $0.96$ & $1.04$\tabularnewline
\hline
\end{tabular}

\caption{SCGB calculations examples (all the methods) for a point on a cylinder
axis ($R$ is the radius of the cylinder) and in the center of a cube
of the size $a$.\label{tab:SCGB-calculations-examples} }

\end{table}

\begin{figure}
\includegraphics[width=0.9\columnwidth]{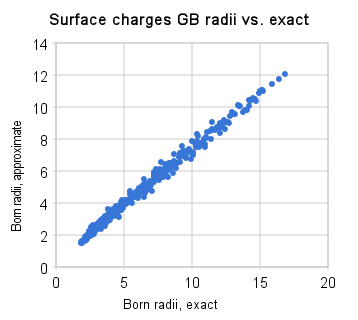}

\caption{SCGB(3) Born radii for the atoms of $2ht7$ protein vs. the {}``exact''
values obtained using a calculation based on surface FEM method.\label{fig:SCGB-Born-radii of 2ht7 protein} }

\end{figure}

All the calculations presented in this Section so far may be fair
but concern only a few oversimplified examples produced for model
systems with idealized geometries. To judge on the actual performance
of the method we turn to a practically interesting realistic system:
solvation energy calculations for $N8$-neuraminidase protein (pdb
accession code $2ht7$). The molecule is composed of $387$ amino
acids and, after all the hydrogen atoms added, has $5866$ atoms.
The results of the calculations are represented on Fig. \ref{fig:SCGB-Born-radii of 2ht7 protein}.
The horizontal axis represents the Born radii taken from {}``exact''
solvation energy $E_{S}$ using the {}``definition'' \[
R_{B}=-q^{2}/2E_{S}.\]
The quantity $E_{S}$ was found {}``exactly'' by solving FEM version
of Eq. (\ref{eq: Eq for sigma}). The vertical axis shows the Born
radii obtained by our SCGB(3) model with the help of (\ref{eq: Surface Born for different alpha})
with $\alpha=3$. The results of the calculations agree pretty well
in the small Born radii region and diverge in the protein center (large
Born radii region). This behavior is well expected, since it is exactly
the center of a large molecule which is the region where the divergence
between the SCGB and exact solvation energy is the most (compare e.g.
with Fig. \ref{fig:Ratios-the-results}).

\begin{figure}
\includegraphics[width=0.9\columnwidth]{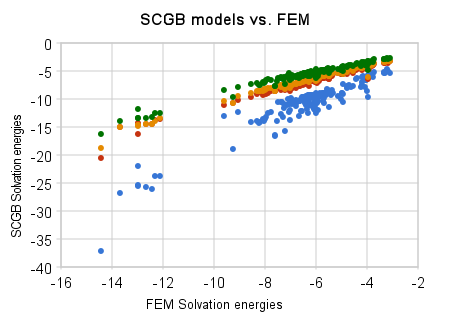}

\caption{$580$ proteins, SCGB solvation vs. the surface FEM method.\label{fig:580proteins-SCGB}}

\end{figure}

The single-charge calculations represented on Fig. \ref{fig:SCGB-Born-radii of 2ht7 protein}
is an interesting but not an ultimate test of the model. What counts
in realistic approximations is of course the solvation energy of a
large molecule with a complicated shape of the molecular surface and
a sophisticated atomic charge distribution. We applied all the four
SCGB models to a set of $580$ proteins from the Quantum Pharmaceuticals
Binding library and presented the results (the data on the vertical
axis) in correlation with the solvation energy obtained with a surface
FEM method (the data on the horizontal axis) on Figure \ref{fig:580proteins-SCGB}.
The blue dots show the performance of SCGB with the Born radii obtained
with the standard CFA formulas. The CFA-based method fails pretty
miserably, whereas all the other SCGB are in good agreement with the
{}``exact'' FEM calculations. Although all three post-CFA SCGB methods
are nearly all as good as each other, the green dots representing
the $SCGB(4)$ model give a somewhat better approximation.

\begin{figure}
\includegraphics[width=0.9\columnwidth]{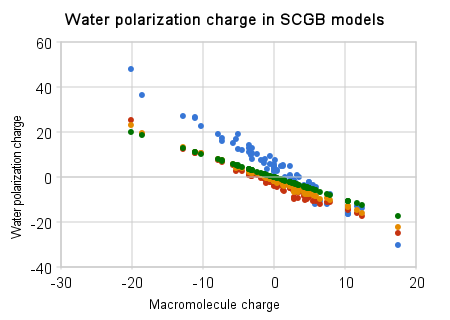}

\caption{Water polarization charge compensation (overall neutrality) demonstration
SCGB calculations.\label{fig:Water-polarization-charge}}

\end{figure}
The reason behind the distinction of the $SCGB(4)$ model may stem
from the {}``inherent'' absolute neutrality of the system of the
protein and the water polarization charges in the model. Indeed, Eq.
(\ref{eq: Sigma_S in FSBE approximation}) for the surface charges
density combined with Eq. (\ref{eq: Surface Born for different alpha})
with $\alpha=3$ give ensure overall neutrality of the system \begin{equation}
Q_{S}=\int df^{\prime}\sigma_{S}\left(\mathbf{r}^{\prime}\right)=-\sum_{j}q_{j}\label{eq: Total surface charge}\end{equation}
for arbitrary surface geometry and the charges distributions. This
does not tell of course that the other two SCGB models are much worse,
the abilities of the methods to recover correct solvation energies
are approximately the same. 

Few concluding remarks should be placed here. Obviously SCGB is not
able to provide exact solutions to the electrostatics problem. In
fact in many of the practical applications this may well not be an
issue: genuine water environment is neither continuous or describable
in terms of simple electrostatics. SCGB is clearly computationally
superior to classic GB implementations both in speed and accuracy
since the calculations can be done in $O(N)$ instead of $O(N^{2})$.
In fact, the real comparison should be made to iterative surface electrostatic
solvers, which can also be made $O(N)-$fast. The advantage comes
from the fact that SCGB solution can be obtained in a number of steps
roughly equal to the number of operations required for a single iteration
of surface based FEM electrostatics solver. Another advantage of SCGB
stems from availability of numerical derivatives for any surface implementation
with surface areas and normals.

The authors are indebted to Quantum Pharmaceuticals for support. The
solvation energy contribution introduced this report is implemented
in a number of Quantum Pharmaceuticals models and employed in Quantum's
drug discovery applications. PCT application is filed.

\bibliographystyle{apsrev}
\bibliography{../Qrefs}

\end{document}